\documentclass[aip,pop,superscriptaddress,reprint,floatfix]{revtex4-1}

\usepackage[plainpages=false, colorlinks=true, anchorcolor=blue, linkcolor=blue, citecolor=blue, bookmarks=false]{hyperref}

\usepackage{amsmath, amsthm, amssymb}
\usepackage{graphicx}

\newcommand{\pd}[2]{\frac{\partial #1}{\partial #2}}

\newcommand{\DS}{\displaystyle}

\begin{document}

\preprint{AIP/123-QED}

\title[Numerical modeling of laser-driven experiments aiming to demonstrate turbulent dynamo]
{Numerical modeling of laser-driven experiments aiming to demonstrate magnetic field amplification via
turbulent dynamo}

\author{P.~Tzeferacos}
\email{petros.tzeferacos@flash.uchicago.edu}
\affiliation{Department of Astronomy and Astrophysics, University of Chicago, Chicago, IL 60637, USA}
\affiliation{Department of Physics, University of Oxford, Oxford OX1 3PU, UK}

\author{A.~Rigby} 
\affiliation{Department of Physics, University of Oxford, Oxford OX1 3PU, UK}

\author{A.~Bott}
\affiliation{Department of Physics, University of Oxford, Oxford OX1 3PU, UK}

\author{A.~R.~Bell}
\affiliation{Department of Physics, University of Oxford, Oxford OX1 3PU, UK}

\author{R.~Bingham}
\affiliation{Rutherford Appleton Laboratory, Chilton, Didcot OX11 0QX, UK}
\affiliation{Department of Physics, University of Strathclyde, Glasgow G4 0NG, UK}

\author{A.~Casner}
\affiliation{CEA, DAM, DIF, F-91297 Arpajon, France}

\author{F.~Cattaneo}
\affiliation{Department of Astronomy and Astrophysics, University of Chicago, Chicago, IL 60637, USA}

\author{E.~M.~Churazov}
\affiliation{Max Planck Institute for Astrophysics, D-85741 Garching, Germany}
\affiliation{Space Research Institute (IKI), Moscow, 117997, Russia}

\author{J.~Emig}
\affiliation{Lawrence Livermore National Laboratory, Livermore, CA 94550, USA}

\author{N.~Flocke}
\affiliation{Department of Astronomy and Astrophysics, University of Chicago, Chicago, IL 60637, USA}

\author{F.~Fiuza}
\affiliation{SLAC National Accelerator Laboratory, Menlo Park, CA 94025, USA}

\author{C.~B.~Forest}
\affiliation{Physics Department, University of Wisconsin-Madison, Madison, WI 53706, USA}

\author{J.~Foster}
\affiliation{AWE, Aldermaston, Reading, West Berkshire, RG7 4PR, UK}

\author{C.~Graziani}
\affiliation{Department of Astronomy and Astrophysics, University of Chicago, Chicago, IL 60637, USA}

\author{J.~Katz}
\affiliation{Laboratory for Laser Energetics, University of Rochester, Rochester, NY 14623, USA}

\author{M.~Koenig}
\affiliation{Laboratoire pour l'Utilisation de Lasers Intenses, UMR7605, CNRS CEA, Universit\'e Paris VI Ecole Polytechnique, France}

\author{C.-K.~Li}
\affiliation{Massachusetts Institute of Technology, Cambridge, Massachusetts 02139, USA}

\author{J.~Meinecke}
\affiliation{Department of Physics, University of Oxford, Oxford OX1 3PU, UK}

\author{R.~Petrasso}
\affiliation{Massachusetts Institute of Technology, Cambridge, Massachusetts 02139, USA}

\author{H.-S.~Park}
\affiliation{Lawrence Livermore National Laboratory, Livermore, CA 94550, USA}

\author{B.~A.~Remington}
\affiliation{Lawrence Livermore National Laboratory, Livermore, CA 94550, USA}

\author{J.~S.~Ross}
\affiliation{Lawrence Livermore National Laboratory, Livermore, CA 94550, USA}

\author{D.~Ryu}
\affiliation{Department of Physics, UNIST, Ulsan 689-798, Korea}

\author{D.~Ryutov}
\affiliation{Lawrence Livermore National Laboratory, Livermore, CA 94550, USA}

\author{K.~Weide}
\affiliation{Department of Astronomy and Astrophysics, University of Chicago, Chicago, IL 60637, USA}

\author{T.~G.~White}
\affiliation{Department of Physics, University of Oxford, Oxford OX1 3PU, UK}

\author{B.~Reville}
\affiliation{School of Mathematics and Physics, Queens University Belfast, Belfast BT7 1NN, UK}

\author{F.~Miniati}
\affiliation{Department of Physics, ETH Z\"urich, CH-8093 Z\"urich, Switzerland}

\author{A.~A.~Schekochihin}
\affiliation{Department of Physics, University of Oxford, Oxford OX1 3PU, UK}

\author{D.~H.~Froula}
\affiliation{Laboratory for Laser Energetics, University of Rochester, Rochester, NY 14623, USA}

\author{G.~Gregori}
\affiliation{Department of Physics, University of Oxford, Oxford OX1 3PU, UK}
\affiliation{Department of Astronomy and Astrophysics, University of Chicago, Chicago, IL 60637, USA}

\author{D.~Q.~Lamb}
\affiliation{Department of Astronomy and Astrophysics, University of Chicago, Chicago, IL 60637, USA}

\date{\today}% It is always \today, today,

\begin{abstract}
The universe is permeated by magnetic fields, with strengths ranging from a femtogauss in the voids 
between the filaments of galaxy clusters to several teragauss in black holes and neutron stars.  The standard model behind
cosmological magnetic fields is the nonlinear amplification of seed fields via turbulent dynamo 
to the values observed.
We have conceived experiments that aim to demonstrate and study the turbulent 
dynamo mechanism in the laboratory.  Here we describe the design of these experiments through simulation
campaigns using FLASH, a highly capable radiation magnetohydrodynamics code that we have 
developed, and large-scale three-dimensional simulations on the {\em Mira} supercomputer at Argonne National Laboratory.
The simulation results indicate that the experimental platform may be capable of reaching a turbulent plasma
state and study dynamo amplification. We validate and compare our numerical results with a small
subset of experimental data using synthetic diagnostics.
\end{abstract}

%                             % Classification Scheme.
%Use showkeys class option if keyword
\maketitle

\section{Introduction}
\label{label:Introduction}

Magnetic fields are encountered throughout the universe\cite{parker1979}. Observational methods based on Faraday rotation
and polarization measurements, Zeeman effect, magneto-bremsstrahlung, even \emph{in situ} measurements
in the case of proximal astrophysical objects, have revealed the broad range of values of cosmical magnetic fields\cite{Zeldovich1983}:
from a femtogauss in the tenuous voids between galaxy cluster filaments, to several microgauss in galaxies
and galaxy clusters, milligauss in molecular clouds,  a few gauss in planets, tens of kilogauss in ordinary stars and accretion disks,
megagauss in white dwarfs, and many teragauss in the vicinity of black holes and neutron stars.
Astrophysical fields are often ``strong,'' in the sense that their energy can amount to a substantial fraction of system's
energy budget, making them salient agents in astrophysical and cosmological phenomena.
This, in conjunction with their ubiquity, has led naturally to the two-fold question of their origin: (1) how
are magnetic fields generated and (2) how do they reach and maintain such large values?

The answer to this question is commonly expressed in terms of dynamo action that operates on seed magnetic
fields\cite{Moffatt1978, parker1979, Kulsrud1997}. Cosmological seed magnetic fields can be generated via a number
of mechanisms, such as plasma instabilities and thermal electromotive forces\cite{Zeldovich1983}, the Biermann
battery effect\cite{Biermann1950} that arises from misaligned electron pressure and density gradients, or the
Weibel instability\cite{Weibel1959} that can occur in collisionless shocks\cite{schlickeiser2003}. These seed fields are
then amplified by the hydromagnetic dynamo mechanism, which achieves a sustained conversion of kinetic energy into magnetic
energy throughout the bulk of an electrically conducting fluid. This mechanism was first invoked  almost a century ago
for solar magnetic fields\cite{Larmor1919}.

An attractive feature of dynamos is that the requirements for their operation are modest.
The two key ingredients are fluid motions that are not too symmetric\cite{Cowling1933,Braginsky1964} and high
electrical conductivity\cite{Elsasser1946,Backus1958}. Both of these requirements are amply satisfied by the
turbulent motions and high magnetic Reynolds numbers prevalent in most astrophysical
situations\cite{Zeldovich1983}, supporting the expectation that dynamo action is widespread in astrophysics\cite{parker1979,Krause1980,Zeldovich1983}.
While astrophysical dynamos come in many flavors\cite{Tobias2013}, they are often distinguished\cite{schekochihin2007} between large-scale
(or mean-field) dynamos, in which the magnetic field grows at scales larger
than those of the fluid motion, and small-scale (or fluctuation) dynamos, where the growth occurs at
or below the outer scales of motion. In this article we will concern ourselves with small-scale dynamo,
at magnetic Prandtl numbers (i.e., magnetic-to-fluid Reynolds number ratio) smaller than unity\cite{schekochihin2007}.
Astrophysical environments with small magnetic Prandtl numbers include planetary cores,
stellar convection zones, the galactic disk, and parts of the interstellar medium\cite{Zeldovich1983}.

Even though conditions favorable for dynamos are common in astrophysics, they are
extremely difficult to realize in laboratory experiments\cite{Verhille2010}. Thus, so far, our physical
intuition in the workings of dynamos is mostly based on theoretical considerations
and numerical modeling\cite{Biskamp2003,Brandenburg2005,Lazarian2006,Kulsrud2008,Brandenburg2011,Tobias2013}.
The reasons for this state of affairs can be easily explained. 
The two natural working fluids for laboratory dynamo experiments are liquid
metals\cite{Gailitis2000, Gailitis2001, Monchaux2007} and strongly ionized gases, i.e., plasmas.
The electrical conductivity of liquid metals, however, makes reaching high magnetic
Reynolds numbers difficult. Conversely, hot plasmas are much better electrical conductors -- therefore capable
of reaching high magnetic Reynolds numbers -- but they tend to be magnetically confined in fusion
devices\cite{Bellan2000} with gas-to-magnetic pressure ratios  $\beta \ll 1$, 
therefore unsuitable to study how they became strongly magnetized in the first place.
Ideally, the aim should be to produce an initially low-magnetization plasma at high
magnetic Reynolds numbers that can, in principle, be used to study dynamo action
in the laboratory. This approach, if successful could provide a much-needed
experimental component to the study of dynamos.

The advent of high-power lasers has opened a new field of research where, using simple scaling
relations \cite{ryutov1999,ryutov2000}, astrophysical environments can be reproduced in the
laboratory \cite{remington1999,Gregori2015}. The similarity achieved is sufficiently close
to make such experiments relevant and informative, in terms of enabling the demonstration
and study of the fundamental physical processes in play.

We have conceived experiments that aim to achieve turbulent dynamo in the laboratory.
The results of these experiments are discussed in a companion paper\cite{Tzeferacos2017}.
In this paper, we describe the design of the experiments through simulation campaigns using FLASH, a
radiation-magnetohydrodynamics (MHD) code that we have developed and large-scale three-dimensional simulations on
the {\em Mira} supecomputer at Argonne National Laboratory (ANL), and the validation of these simulations using
a subset of the experimental data. Three-dimensional simulations were required in order to represent with high fidelity both the geometry 
of the targets and key physical processes, so as to be predictive.  The simulations were vital to ensuring
that the experiments achieved the strong turbulence and large magnetic Reynolds numbers needed for turbulent
dynamo to operate. The simulations
were also necessary to determine when to fire the diagnostics, since the experiments last tens of nanoseconds but the
strongly amplified magnetic fields persisted for only a fraction of this time.

In \S II, we describe the High Energy Density Laboratory Plasma (HEDLP) capabilities of the FLASH code
that were used in the simulations we performed.
In \S III, we discuss key elements of the platforms we used in previous experiments.  These
platforms informed the design of the experiments that we describe here.
In \S IV, we describe the simulations that we performed and that led to the fielded experimental platform, as well as the final design.
In \S V, we discuss the simulation results, as well as their validation against a subset of experimental data.

\section{Simulation code}
\label{label:simulation_code}

We use the FLASH code \cite{Fryxell2000,dubey2009} to carry out the large-scale simulations of our
laser experiments to study the origin of cosmic magnetic fields. FLASH is a publicly
available\footnote{The FLASH code is available at http://flash.uchicago.edu}, parallel,
multi-physics, adaptive mesh refinement (AMR), finite-volume Eulerian hydrodynamics and
MHD code. FLASH scales well to over 100,000 processors, and uses a variety of parallelization
techniques including domain decomposition, mesh replication, and threading to make optimal use of hardware resources.

Extensive HEDLP capabilities\cite{tzeferacos2015} have been added to FLASH, making it a suitable code for
simulating laser-driven plasma experiments. The system of partial differential equations employed in the numerical
modeling of the experiment has the general form  
\begin{equation}
\DS \pd{{\bf U}}{t} + \nabla \cdot {\bf F}({\bf U}) =  {\bf S}({\bf U}),
\end{equation}
where ${\bf U}$ denotes the conserved variables (e.g. ${\bf U} \equiv (\rho,\,{\bf m},\,{\bf B},\,{\cal E})^T$
for ideal MHD), 
${\bf F}({\bf U})$ the fluxes and ${\bf S}({\bf U})$ the source terms. Here we use
the customary notation for density ($\rho$), momentum density (${\bf m}$),
magnetic field (${\bf B}$) and total energy density (${\cal E}$).

A single-temperature ideal MHD formulation is insufficient to model HEDLP experiments:
thermal equilibrium between electrons, ions, and radiation is disrupted by a number of
physical processes and equilibration times can be sufficiently long to warrant a multi-temperature
treatment. To accomplish this, we extended the ideal MHD system of equations by retaining a
single-fluid treatment while considering different temperatures for one species of ions, electrons, and radiation
(i.e. three temperatures or 3T). This extension requires that the total pressure be defined as
$\DS p_{tot} = p_i +p_e + p_r + {\bf B}^2/2 $,
where the subscripts $i$, $e$, and $r$ denote ions, electrons, and radiation, respectively.
The continuity and momentum equations are given by 
\begin{equation}
\label{eq:continuity}
\frac{\partial {\rho}}{\partial t} + \nabla \cdot (\rho {\bf u}) = 0,
\end{equation}
\begin{equation}
\label{eq:momentum}
\frac{\partial {\rho {\bf u} }}{\partial t} + \nabla \cdot \left[ \rho {\bf u u} + p_{tot}{\bf I} - {\bf B B}\right] = 0.
\end{equation}
In the induction equation, we consider the generalized form of Ohm's law
$\DS {\bf E} = -{\bf u} \times {\bf B} +\eta {\bf J} - \nabla p_e /(q_e n_e)$,
where $q_e$ is the electron charge, $n_e$ the electron number density,
$\eta$ the magnetic resistivity\cite{braginskii1965}, ${\bf J} = \nabla \times {\bf B}$ the current density, 
and ${\bf E}$ the electric field. Note that, in our isotropic treatment we do not include the Nernst term\cite{Haines1986}, because its
effect is not important for the plasma conditions described here; it is however relevant in many other laser-based experiments.
The induction equation then reads
\begin{equation}
\label{eq:induction}
\frac{\partial {{\bf B} }}{\partial t} + \nabla \times \left[ - {\bf u} \times {\bf B} \right] + \nabla \times \left[ \eta {\bf J} -  \frac{\nabla p_e}{q_e n_e} \right] = 0,
\end{equation}
where the non-ideal terms on the left-hand side include magnetic diffusivity and the
Biermann battery term\cite{fatenejad2013a,graziani2015}. We also evolve the total energy density equation,
\begin{eqnarray}
\label{eq:energy}
\frac{\partial {\cal E}}{\partial t} + &&\nabla \cdot \left[\left({\cal E} + p_{tot}\right){\bf u}
- \left({\bf u}\cdot{\bf B}\right){\bf B}\right] \nonumber\\
-&& \nabla \cdot \left[{\bf B} \times \left(\eta {\bf J} - \frac{\nabla p_e}{q_e n_e}\right)  \right]
= - \nabla \cdot {\bf q}  + S,
\end{eqnarray}
where the total energy density is given by $\DS {\cal E}=\rho E_{tot}=\rho e_{int} +\rho{\bf u}^2/2 + {\bf B}^2/2$
and the total specific internal energy includes the radiation energy, $e_{int} = e_i + e_e + e_r$.
The total heat flux $\DS {\bf q}= {\bf q_e}+{\bf q_r}$ is the sum of the electron heat flux $\DS {\bf q_e}=-\kappa \nabla T_e$
and the radiation flux ${\bf q_r}$. For the former, we denote with $\kappa$ the electron conductivity\cite{Spitzer1962},
and $T_e$ the electron temperature. The source term $\DS S$ encompasses external contributions of energy,
typically due to laser heating.

To treat the 3T components we also consider the non-conservative energy equations
for electrons, ions, and radiation. These can be written as
\begin{eqnarray}
\label{eq:e_ion}
\frac{\partial \rho e_i}{\partial t} + && \nabla \cdot \left(\rho e_i {\bf u}\right) + p_i\nabla\cdot {\bf u} =\rho \frac{c_{v,e}}{\tau_{ei}}  \left(T_e - T_i \right),\\
\label{eq:e_ele}
\frac{\partial \rho e_e}{\partial t} + && \nabla \cdot \left(\rho e_e {\bf u}\right) + p_e\nabla\cdot {\bf u} =\rho \frac{c_{v,e}}{\tau_{ei}}  \left(T_i - T_e \right)\\
- && \nabla \cdot {\bf q}_e \nonumber + Q_{abs} - Q_{emis} + Q_{las} +Q_{Ohm},\\
\label{eq:e_rad}
\frac{\partial \rho e_r}{\partial t} + && \nabla \cdot \left(\rho e_r {\bf u}\right) + p_r\nabla\cdot {\bf u} =- \nabla \cdot {\bf q}_r \\
- && Q_{abs} + Q_{emis},\nonumber
\end{eqnarray}
where $c_{v,e}$ is the electron specific heat, $\tau_{ei}$ the ion-electron relaxation time,
$Q_{abs}$ the rate of increase of the electron internal energy density due to radiation
absorption, $Q_{emis}$ the rate of decrease due to radiation emission, and $Q_{Ohm}$ the rate of increase
due to Ohmic heating. The system closes with 3T equations of
state (EoS) that connect internal energies, temperatures, and pressures of the components. This is
accomplished using either an analytical prescription or, more frequently, through tabulated EoS.

The system of equations (\ref{eq:continuity}-\ref{eq:energy}) is a mixed hyperbolic-parabolic
system. All the terms on the right-hand side of the equations are operator-split from the solution
of the non-ideal single-fluid magneto-hydrodynamics.
The latter is handled using the single-step, time marching algorithm of the unsplit staggered
mesh (USM) \cite{Lee2009, Lee2013} for Cartesian coordinates and its extension to cylindrical systems
\cite{Tzeferacos2012}. Both resistivity and the Biermann battery term are included in the
staggered electric field, which allows us to preserve magnetic field solenoidality at
machine accuracy through constrained transport.

In order to utilize 3T EoS and properly distribute the update of $e_{int}$ to its
components, we advance the auxiliary equations (\ref{eq:e_ion}-\ref{eq:e_rad}).
However, the work terms $\DS p_{s}\nabla {\bf u}$  (with $s$ denoting
ions, electrons, or radiation) are ill-defined at shocks. To overcome this,
we employ a method inspired by the radiation-hydrodynamics code RAGE
\cite{Gittings2008}, which distributes the change due to work and total
shock-heating, recovered from the solution of equations (\ref{eq:continuity}-\ref{eq:energy}),
based on the pressure ratio of the components.

The right-hand side of equations (\ref{eq:e_ion}-\ref{eq:e_rad})
is in turn operator-split and each physical process is handled separately\cite{tzeferacos2015}.
The physical processes represented in our formulation include 
energy exchange between ions and electrons through collisions,
electron thermal conduction, and radiation transport in the multi-group,
flux-limited diffusion approximation. The last two are solved implicitly using
the HYPRE\footnote{The HYPRE library is available at https://computation-rnd.llnl.gov/linear\_solvers/software.php} library
to retain large time steps. To model the laser heating, we utilize 
laser beams in the geometric optics approximation. These are comprised of rays
whose paths are traced\cite{Kaiser2000} through the computational domain, based on the local refractive
index of each cell. The laser power is deposited at the inverse
bremsstrahlung rate, which depends on local electron number density gradients and
local electron temperature gradients.
The HEDLP capabilities of FLASH have been recently exercised in a number of
experiments\cite{falk2014, yurchak2014, meinecke2014, meinecke2015,Li2016}, as well as in the
experiments described in what follows.
\begin{figure}
\includegraphics[width=\columnwidth]{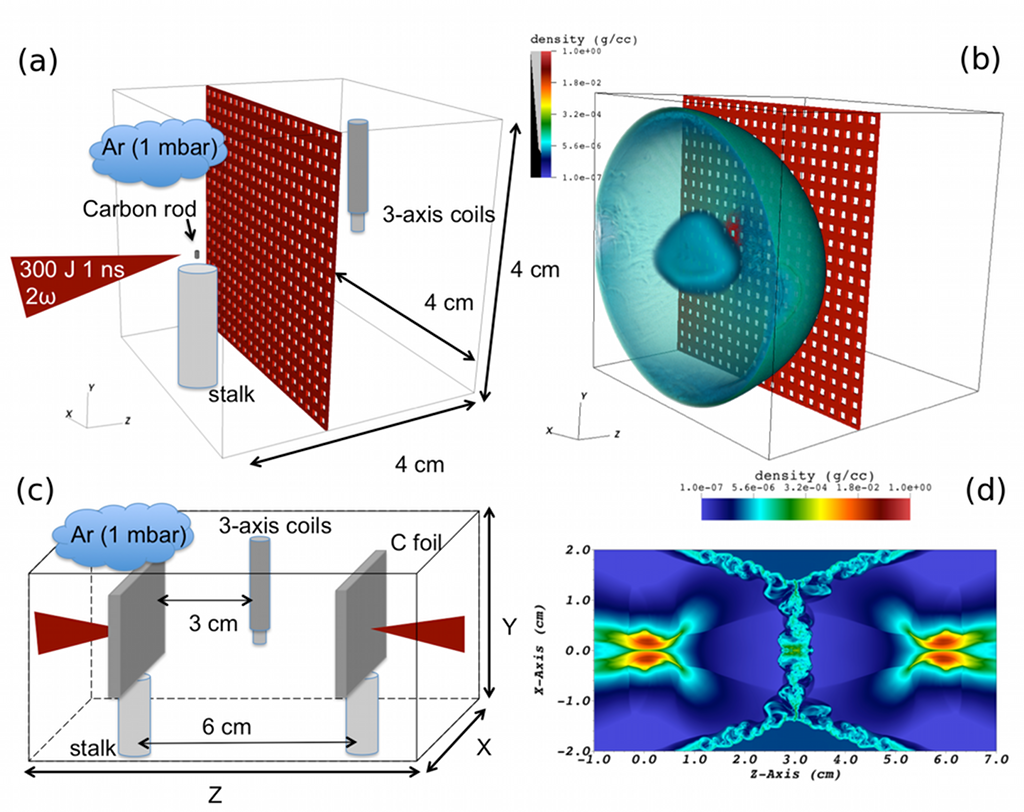}
 \caption{Schematics and numerical simulations of previous experiments conducted at Vulcan. (a) Cartoon of the rod-grid experiment\cite{meinecke2014}. (b)
 3D FLASH simulation of the rod-grid experiment. Displayed is the density logarithm when the shock traverses the plastic grid, stirring turbulence
 that amplifies the Biermann battery generated field by a factor of two\cite{meinecke2014}. Numerical models of this experiment\cite{tzeferacos2015}
 enabled the interpretation of the experimental results. (c) Cartoon of the colliding flows experiment\cite{meinecke2015}, where higher Rm values where
 obtained. (d) 2D cylindrical FLASH simulation of the colliding flows experiment\cite{meinecke2015}.}
\label{fig:vulcan}
\end{figure}

\section{Previous experiments}
\label{label:Vulcan}

We have developed an experimental program to exploit the similarity in scaled laboratory experiments.
In a first set of experiments done using the LULI2000 laser at the Laboratoire
d'Utilisation des Lasers Intenses in France, we successfully demonstrated the creation of seed
magnetic fields at asymmetric shocks \cite{Gregori2012} by the Biermann battery effect \cite{Biermann1950}, as predicted by protogalactic
structure formation simulations \cite{Kulsrud1997}. In these experiments, a carbon rod target was placed in a
chamber filled with helium gas. Laser beams were focused on the target, vaporizing part of it and launching an asymmetric shock into the gas. The seed magnetic fields created at the
shock by the Biermann battery effect were measured using three-axis induction coils.

In a second set of experiments done using the Vulcan laser at the Rutherford-Appleton Laboratory in
the UK, we showed that seed magnetic fields can be amplified in plasmas by the turbulence produced
as shocks interact with strong density inhomogeneities\cite{meinecke2014}, reminiscent of what is observed in the supernova remnant
Cassiopeia A \cite{vandenberg1970}. In these experiments, a carbon rod target was placed in a chamber filled with
argon gas. Laser beams were again focused on the target, vaporizing part of the target and launching
an asymmetric shock into the gas. The interaction of a shock with large density perturbations was
reproduced in the laboratory by passing the shock through a plastic mesh (see Fig. \ref{fig:vulcan}, panels a and b). In this
case, the seed magnetic fields produced by the Biermann battery effect were amplified by the turbulence
produced when the shock passed through the grid. The amplified magnetic field was measured
using three-axis induction coils. Due to the relatively small electron temperatures, the plasma was characterized
by large magnetic resistivity and magnetic Reynolds numbers $\textrm{Rm}\sim1$; as a result, the field was
amplified due to tangling and the magnetic energy followed a $k^{-11/3}$ Golitsyn power law,
a consequence of balancing field advection and resistive diffusion\cite{golitsyn1960,schekochihin2007}.

In a third set of experiments done using the Vulcan laser at the Rutherford-Appleton
Laboratory in the UK, we demonstrated the ability to achieve developed turbulence and the higher
magnetic Reynolds numbers (i.e., the higher velocities and temperatures) needed to produce greater
amplification of seed magnetic fields, a precursor to turbulent dynamo\cite{meinecke2015}. 
In these experiments, lasers were focused on two foil targets in a chamber filled with argon gas, producing plasma jets that collided in the center.
(see Fig. \ref{fig:vulcan}, panels c and d). The collision of the two jets produced developed turbulence in the interaction region that
amplified the seed magnetic fields created by the Biermann battery effect.
However, the Rm values that were obtained ($\sim 10$) were still small for dynamo action\cite{Schekochihin2004,schekochihin2007}.  

Building on these results, we have conceived and designed an experimental platform for the Omega laser facility
at the Laboratory for Laser Energetics at the University of Rochester, in order to demonstrate and study the turbulent dynamo 
mechanism. While the platform combines key elements of our previous experiments to generate turbulent 
plasmas and modest amplifications of seed magnetic fields, it differs in one crucial aspect: according to the simulations,
it may be possible to reach high enough Rm values for turbulent dynamo to operate.

\begin{figure}
\includegraphics[width=\columnwidth]{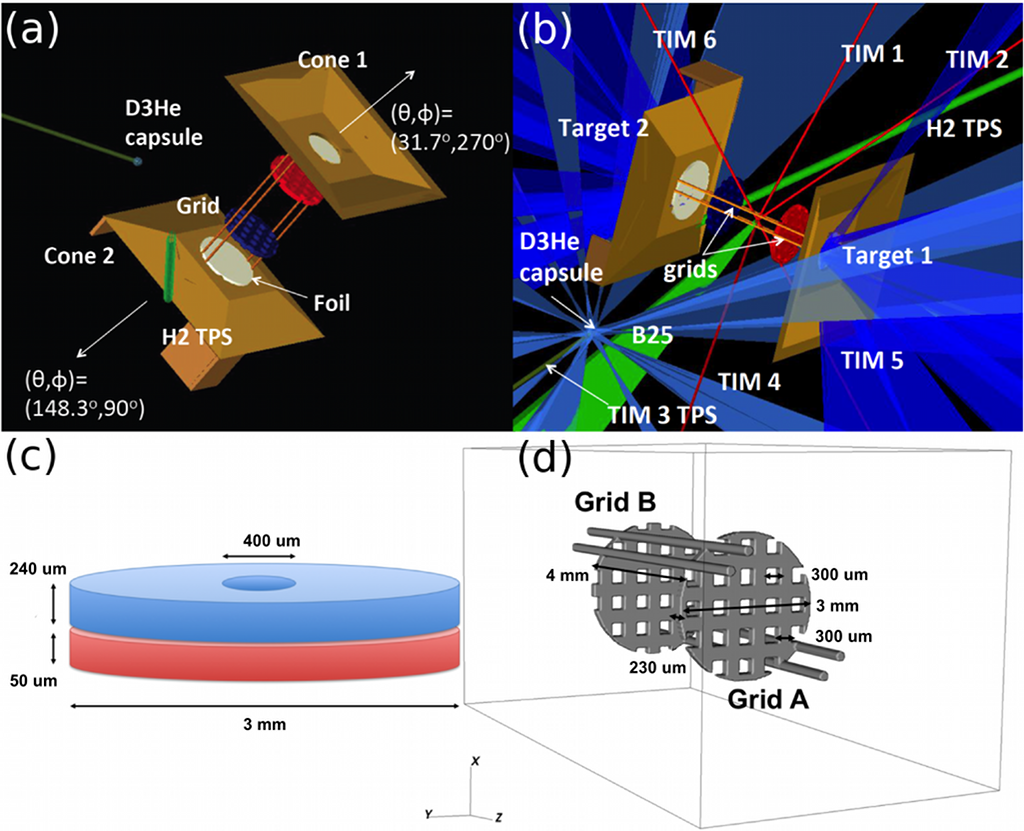}
  \caption{(a) VisRad\footnote{http://www.prism-cs.com/Software/VisRad/VisRad.htm} target configuration, oriented in the Omega chamber. The two
           targets are placed opposite to each other. A pair of grids is situated in the
           propagation path of the flows, while two cones act as shields to both the
           interaction region and the diagnostic instruments. The small D$^3$He capsule
           next to the assembly is the proton source for proton radiography.
           (b) VisRad experimental configuration. The blue beams show the drive
           on the two targets and the D$^3$He capsule. The six TIMs (red lines) show the position
           of the diagnostics. (c) Detail of the composite targets:
           a polystyrene washer with a cylindrical ``well'' is placed on top of a thin chlorine-doped
           polystyrene foil. (d) Design specifications for the polyimide grids, through which the plasma flows will propagate.}
\label{fig:Target}
\end{figure}

\section{Design simulations}
\label{label:design}

\subsection{Target design}
\label{label:target}

To design the experiments, we conducted an extensive series of 2D-cylindrical FLASH radiation-MHD simulations,
followed by a smaller set of 3D FLASH radiation-MHD simulations on the Mira supercomputer at ANL.
The simulations led to an experimental design that combines key elements of each of our two earlier
experiments on Vulcan\cite{meinecke2014,meinecke2015}: a hot plasma flowing through a
grid in the first and two plasma jets colliding in the second. The broad design goals consisted of obtaining  
\begin{itemize}
\item[--] a large kinetic energy reservoir in the turbulent flow to amplify the magnetic
fields to measurable values;
\item[--] large magnetic Reynolds numbers, i.e., high temperatures and velocities, for the turbulent 
dynamo to operate; and
\item[--] sustained turbulence that would persist for a few eddy turnover timescales -- at the driving scale -- so as
to amplify the field to saturation values.
\end{itemize}

In this design, the assembly is comprised of two composite targets and two grids
that are connected by four boron rods (Figure \ref{fig:Target}a).
The composite targets are 3 mm in diameter and consist of a chlorine-doped polystyrene foil, 50 $\mu$m thick,
and a polystyrene washer, 240 $\mu$m thick (Figure \ref{fig:Target}c). The polystyrene washers were
machined so as to have a 400 $\mu$m-diameter cylindrical ``well'' in their centers.  
The two targets are mounted 8 mm apart (the distance measured from the proximate faces of the foils), and
the pair of grids is placed between them. The two grids, made of polyimide, are mounted 4 mm apart -- the distance
is once more measured with respect to their proximate faces -- each of them 2 mm away from the respective
proximate face of the foil-target. The grids have a diameter of 3 mm and a thickness of 230 $\mu$m.
The opening fraction of each grid is 25\%, with 300 $\mu$m-wide holes and a spacing of 300 $\mu$m (Figure \ref{fig:Target}d).
The hole patterns of the grids are offset by 300 $\mu$m with respect to each other,
thus breaking the mirror symmetry of the assembly: grid A has a hole in the center while grid B
does not. Rectangular cones on each target shield the diagnostics from the intense X-ray emission produced when
a sequence of ten 1-ns duration laser beams coming from different angles illuminate each target (Figure \ref{fig:Target}b).

The two targets are driven for either 5 or 10 ns, delivering a total of 5 kJ on an area defined by the laser phase
plates. The radial profile of each beam's circular spot on target can be approximated
by a super-Gaussian of exponent 6 and an e-folding radius of 336 $\mu$m; however, due to variation in the incidence angle, the
illuminated area on each target is the overlap of ten ellipses. The temporal profile of the drive is either a 10 ns ``top-hat''
-- each 1-ns long beam is fired sequentially so as to deliver 500 J per ns -- or a ``staircase'' profile, ramping up the power
towards the end of the drive (500 J/ns for 2 ns, 1000 J/ns for 1 ns, and 1500 J/ns for 2 ns).

The platform described above was designed based on our previous experiments and scores of 2D FLASH cylindrical simulations; many
of its elements reflect the design goals stated at the beginning of this section. The machined washers act as collimators to direct the kinetic energy
of the flows towards the collision region, minimizing lateral expansion; the offset of grids A and B
results in corrugated fronts that will interleave, shear, and trigger Kelvin-Helmholtz instabilities that maximize mixing and the duration of turbulence;
and the thickness of the foil components of the targets was selected so as to achieve large velocities
while avoiding shine-through of the driving lasers, which could disrupt the turbulent flow and generate
strong Biermann battery magnetic fields\cite{Haines1997}.

While 2D simulations can provide useful information in the platform design process, they are not able to
reproduce the experiment with high fidelity. MHD turbulence in two dimensions behaves differently than in
three\cite{Biskamp2000} and,
according to anti-dynamo theorems\cite{Zeldovich1980}, cannot sustain dynamo. Moreover,
the experimental platform has features that break the cylindrical symmetry assumed by our 2D modeling, which can have
significant repercussions on the flow dynamics. Good examples are the square holes of the grids, the presence of the
support rods, and asymmetries in the laser drive -- a consequence of variance in directions and incidence angles of the laser
beam sequences that irradiate the foils. To model the experiment properly, three dimensional simulations are required.

\subsection{Three-dimensional simulations}
\label{label:3d}

\begin{figure}
\includegraphics[width=0.75\columnwidth]{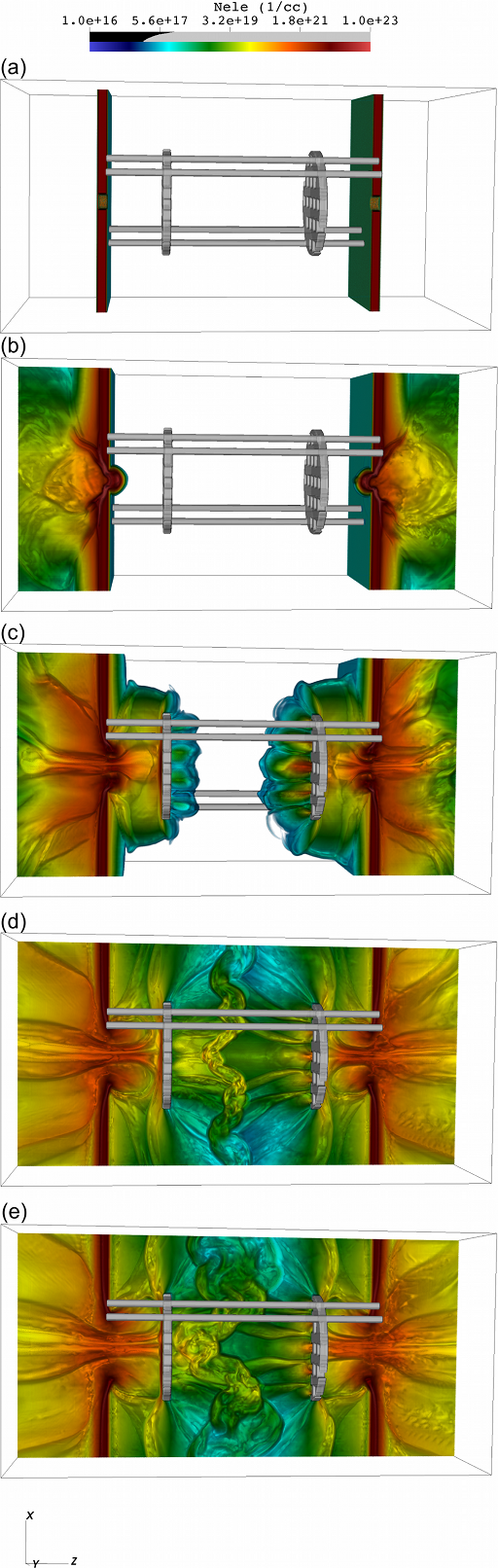}
 \caption{Initial condition and temporal evolution for the 1Cl10ns simulation. (a) Electron number density logarithm (half-rendering)
 and contours of the grids and supporting boron rods at $t=0$ ns. (b) Same as (a) but for $t=8$ ns. (c) Same as (a) but for $t=20$ ns.
 (d) Same as (a) but for $t=35$ ns. (e) Same as (a) but for $t=45$ ns.}
\label{fig:DensEvolution}
\end{figure}

\begin{table}[t!]
\begin{center}
\vspace{-0.3cm}
\caption{Simulation key, target characteristics, and drive.\label{table:Simulations}}
\begin{tabular}{cccc}
\hline\hline
Simulation$^{\textrm{a}}$        & Composition$^{\textrm{b}}$  & Density$^{\textrm{c}}$ & Drive\\
\hline
1Cl10ns                          & C(50.4\%) H(48.3\%) Cl(1\%) &  1.29 g cm$^{-3}$ & 10 ns \\
6Cl10ns                          & C(49.9\%) H(43.8\%) Cl(6\%) &  1.55 g cm$^{-3}$ & 10 ns \\
1Cl5ns                           & C(50.4\%) H(48.3\%) Cl(1\%) &  1.29 g cm$^{-3}$ & 5 ns \\
6Cl5ns                           & C(49.9\%) H(43.8\%) Cl(6\%) &  1.55 g cm$^{-3}$ & 5 ns \\
\hline\hline
\end{tabular}
\end{center}
\vspace{-0.5cm}
\begin{flushleft}
{\footnotesize $^{\textrm{a}}$ The key for each simulation is defined by the chlorine doping percentage and the drive duration.}\\
{\footnotesize $^{\textrm{b}}$ The composition refers only to that of the foil part of the target. The composition of the washer
is that of regular polystyrene (CH).}\\
{\footnotesize $^{\textrm{c}}$ In comparison, the density of the washer is 1.07 g cm$^{-3}$, which highlights the density increase due to the
chlorine doping.}\\
\end{flushleft}
\vspace{-0.8cm}
\end{table}

In this section, we discuss the characteristics of four different 3D FLASH simulations that reflect the majority of the
experimental configurations that we fielded at the Omega laser facility (see also Table \ref{table:Simulations}). The simulations vary
in terms of the material properties of the targets (density and composition of the foils), and the shape and duration of the laser drive.
The initial conditions reflect the design specifications of the platform, discussed in \S \ref{label:target}.
In a computational domain that spans 0.625 cm in $X$ and $Y$, and 1.250 cm in $Z$, we initialize the targets, grids, and rods
that we described in Figure \ref{fig:Target}, at a temperature of $\sim 290$ K. A snapshot of the initial condition for case 1Cl10ns
(logarithm of electron number density and contours of the grids and rods) is shown in Figure \ref{fig:DensEvolution}a.
To simplify the initialization, we omit the diagnostic shields and extend our targets to the domain boundaries, effectively separating
the back of the foils -- where laser illumination occurs -- from the domain center.

The domain is resolved with $\sim 3.3\times10^7$ cells, corresponding to $\sim$~25~$\mu$m per cell width. The boundary conditions
on all sides of the computational box are set to ``outflow'' (zero-gradient), except for the normal component of the magnetic field, which
is recovered through the solenoidality condition. For the multigroup flux-limited radiation diffusion 
we consider 6 energy bins from
0.1 eV to 100 keV. To model accurately the material properties of the chlorinated targets we utilize opacity and EoS tables computed
with PROPACEOS\footnote{PROPACEOS is available at http://www.prism-cs.com/Software/PROPACEOS/PROPACEOS.htm}.
Temporal integration of the non-ideal 3T MHD equations is carried out for $50$ ns, using the second-order unsplit time-marching method of the
USM algorithm\cite{Lee2013}, an extension of the corner transport upwind (CTU) approach\cite{Colella1990}. Spatial reconstruction is done utilizing
the piecewise parabolic method\cite{Colella1984} (PPM) and a minmod limiter. The upwind fluxes are computed with an HLLC\cite{LiHLLC2005}
(Harten-Lax-van Leer Contact) Riemann solver. Implicit solves for radiation and electron thermal conduction are carried out using a conjugate
gradient method (PCG), preconditioned with algebraic multigrid (AMG), as implemented in the HYPRE library. 

To model accurately the laser drive, we implemented the spatial and temporal specifications of each of the twenty Omega driver beams separately.
This was done to ensure that the interplay between obliqueness of incidence angle and target deformation due to the drive would be
captured correctly. Each 3$\omega$ beam is simulated using 16,000 rays per timestep, achieving good statistics and low Poisson noise
in the energy deposition. The $5\times10^{11}$~W power in each beam is distributed assuming the spatial beam profile, mentioned above. 

The temporal evolution of the system is shown in Figure \ref{fig:DensEvolution}. In the simulations, the laser beams ablate the back
of the foil targets and a pair of hot plasma plumes are created and expand outwards. The laser-target interaction generates strong
magnetic fields due to the Biermann battery mechanism\cite{Haines1997}, which are ``flux-frozen'' into and advected by the plasma.
The ablation results in a pair of shocks -- driven inside the chlorinated polystyrene foils -- that break out and propagate supersonically
towards the grids (Figure \ref{fig:DensEvolution}b).
The lateral expansion of the inwards-moving plasma flows is inhibited by the collimating effect of the washers. The laser drive (for this case)
persists for 10 ns and is turned off shortly after the break-out. Subsequently, the flows traverse the grids to form ``finger'' formations
and corrugated fronts of a characteristic length-scale $\cal{L}$ $\sim$ 600 $\mu$m -- the sum of a hole width and a hole spacing --
and continue towards the center of the domain (Figure \ref{fig:DensEvolution}c). The flows then collide to form a cup-shaped
interaction region of hot, subsonic turbulent plasma with an outer scale defined by $\cal{L}$ (Figure \ref{fig:DensEvolution}d).
The bottom of the ``cup'' is pointing towards grid B, a result of $\cal{L}$ being comparable to the thickness of the interaction region:
as grid A has a center hole, the locally increased mass flux from grid A results in the deformation. At late times
(Figure \ref{fig:DensEvolution}e), the interaction region thickens and slowly drifts towards grid B, gradually cooling by advection (primarily)
and radiation.   

\begin{figure}
\includegraphics[width=\columnwidth]{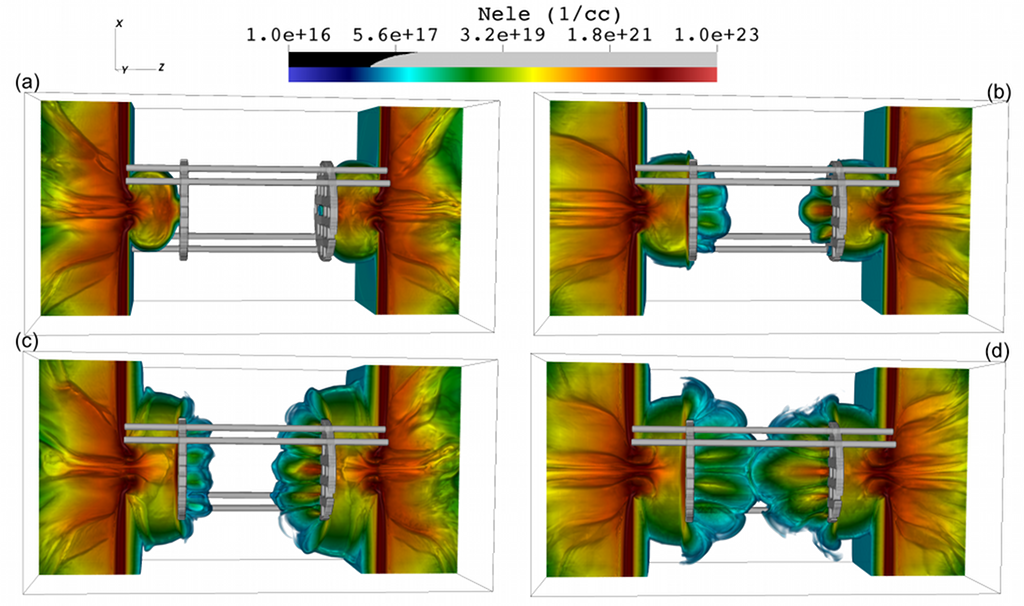}
 \caption{Front positions (electron density logarithm) at 20ns for the various simulated cases, (a) 6Cl10ns, (b) 6Cl5ns, (c) 1Cl10ns, and (d) 1Cl5ns}
\label{fig:timing}
\end{figure}

\begin{figure}
\includegraphics[width=\columnwidth]{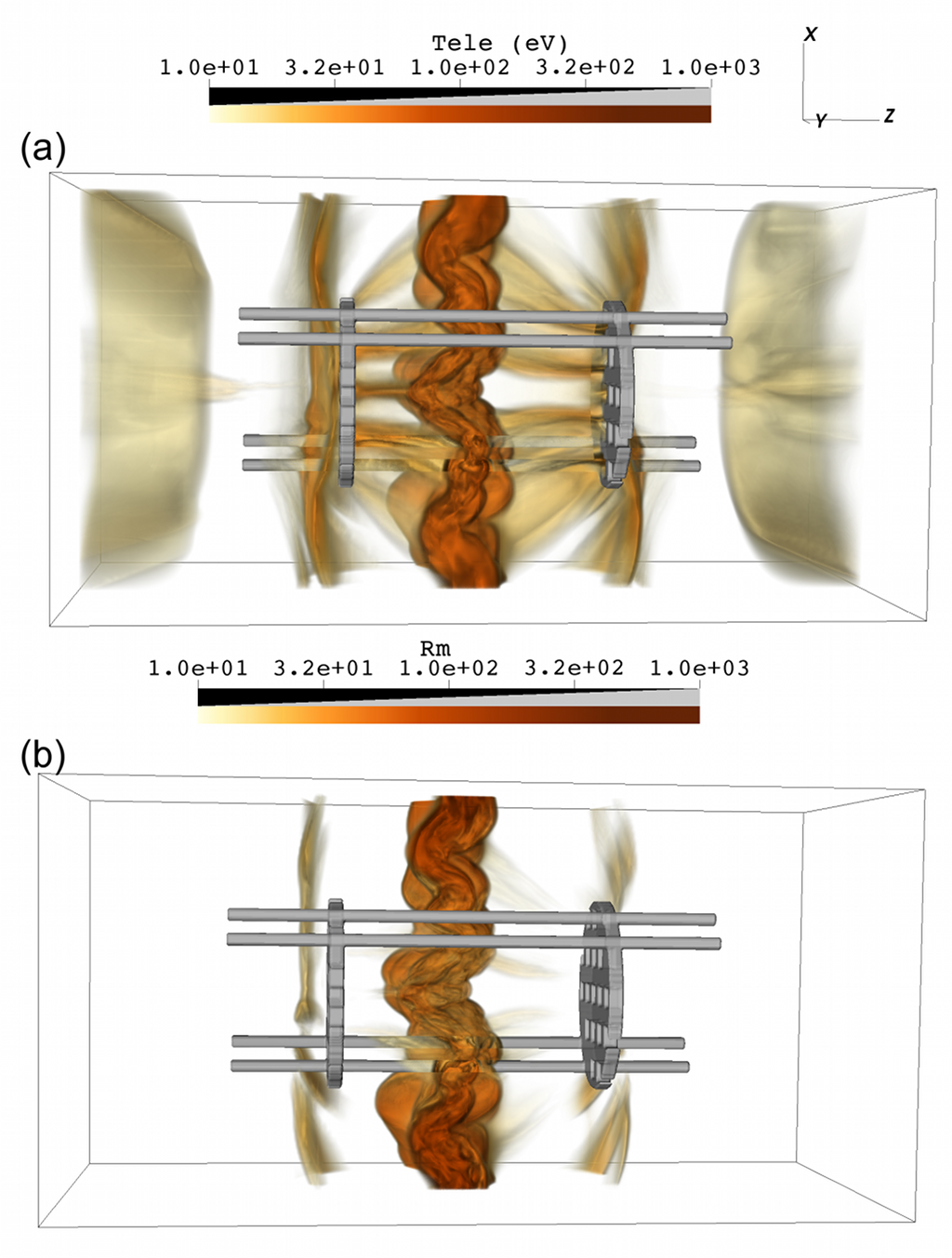}
 \caption{Volume rendering of (a) the electron temperature in eV and (b) the magnetic Reynolds number (at scale $\cal{L}$), for case 1Cl10ns at 35 ns.}
\label{fig:TeleRm}
\end{figure}

\begin{table*}[t!]
\begin{center}
\vspace{-0.3cm}
\caption{Simulated plasma properties for case 1Cl10ns prior to and after collision.\label{table:PlasmaProperties}}
\begin{tabular}{lccc}
\hline\hline
Plasma property     & Formula  & Prior to collision$^{\textrm{a}}$  & After collision$^{\textrm{b}}$\\
\hline   
Electron density $N_{\textrm{ele}}$ ($\textrm{cm}^{-3}$) &  --   & $\sim 1\times10^{18}$ & $\sim 8\times10^{19}$\\
Ion density $N_{\textrm{ion}}$ ($\textrm{cm}^{-3}$)      &  --   & $\sim 4\times10^{17}$ & $\sim 3\times10^{19}$\\
Electron temperature $T_{\textrm{ele}}$ (eV)             &  --   & $\sim 60-90$          & $\sim 150-350$       \\
Ion temperature $T_{\textrm{ion}}$ (eV)                  &  --   & $\sim 100-120$        & $\sim 150-350$       \\
Average ionization $Z$                                     &  --   & $\sim 3.6$            & $\sim 3.6$           \\
Average atomic weight $A$ (a.m.u.)                         &  --   & $\sim 6.8$            & $\sim 6.8$           \\
Flow velocity $u$ ($\textrm{cm\,s}^{-1}$)                  &  --   & $\sim 2.\times10^{7}$ & $ \sim 1.4\times10^{7}$\\
Coulomb logarithm ln$\Lambda$                           &$23.5 - \ln{\left(N_{\textrm{ele}}^{1/2} T_{\textrm{ele}}^{-5/4}\right)}-\sqrt{10^{-5}+\frac{\left(\ln{(T_{\textrm{ele}})}-2\right)^2}{16}}$ & $\sim 7.4-7.8$ & $\sim 6.1-6.9$\\
Sound speed $C_{\textrm{s}}$ ($\textrm{cm\,s}^{-1}$)     &$9.80 \times 10^5 \, \frac{\left[Z T_{\textrm{ele}}+(5/3)T_{\textrm{ion}} \right]^{1/2}}{A^{1/2}}$ & $\sim 7-9\times 10^6$ & $\sim 1.1-1.6\times 10^7$\\
Mach number $M$   &  $u/C_{\textrm{s}}$                    & $\sim 2 - 3$& $\lesssim 1$ \\
Ion-ion mean free path $\lambda_{ii}$ ($\textrm{cm}$)    &  $2.88 \times 10^{13} \, \frac{T_{\textrm{ion}}^2}{Z^4 N_{\textrm{ion}}   \ln{\Lambda}}$   & $\sim5-8\times10^{-4}$& $\sim 0.2-1\times10^{-4}$\\
Magnetic Reynolds number Rm                              &    $u\cal{L}/\eta$ $\left(\eta = 3.2 \times 10^5 \,\textrm{cm}^2\,\textrm{s}^{-1}\,  \frac{Z \ln{\Lambda}}{T_{\textrm{ele}}^{3/2}}\right)$        & $\sim 60 - 120 $& $\sim 300 - 900^{\textrm{c}} $\\
Reynolds number Re                                       &     $u\cal{L}/\nu$ $\left(\nu = 1.92 \times 10^{19} \,\textrm{cm}^2\,\textrm{s}^{-1}\,  \frac{T_{\textrm{ion}}^{5/2}}{A^{1/2} Z^{4} N_{\textrm{ion}}  \ln{\Lambda} }\right)$     & $\sim 540-850$&$\sim 1300-8300$\\
Magnetic Prandtl number Pm                               &      Rm / Re      & $\gtrsim 0.1$ & $\gtrsim 0.1$\\
\hline\hline
\end{tabular}
\end{center}
\vspace{-0.5cm}
\begin{flushleft}
{\footnotesize $^{\textrm{a}}$ Estimated in the tracking control volume, 2 ns prior to collision.}\\
{\footnotesize $^{\textrm{b}}$ Characteristic values in a 10 ns time range after collision.}\\
{\footnotesize $^{\textrm{c}}$ Peak values of $1,300-1,600$ within the first 4 ns after collision, consistent with
threshold estimates for the small Pm regime\cite{schekochihin2007}.}\\
\end{flushleft}
\vspace{-0.8cm}
\end{table*}

\begin{figure}[!h]
\includegraphics[width=\columnwidth]{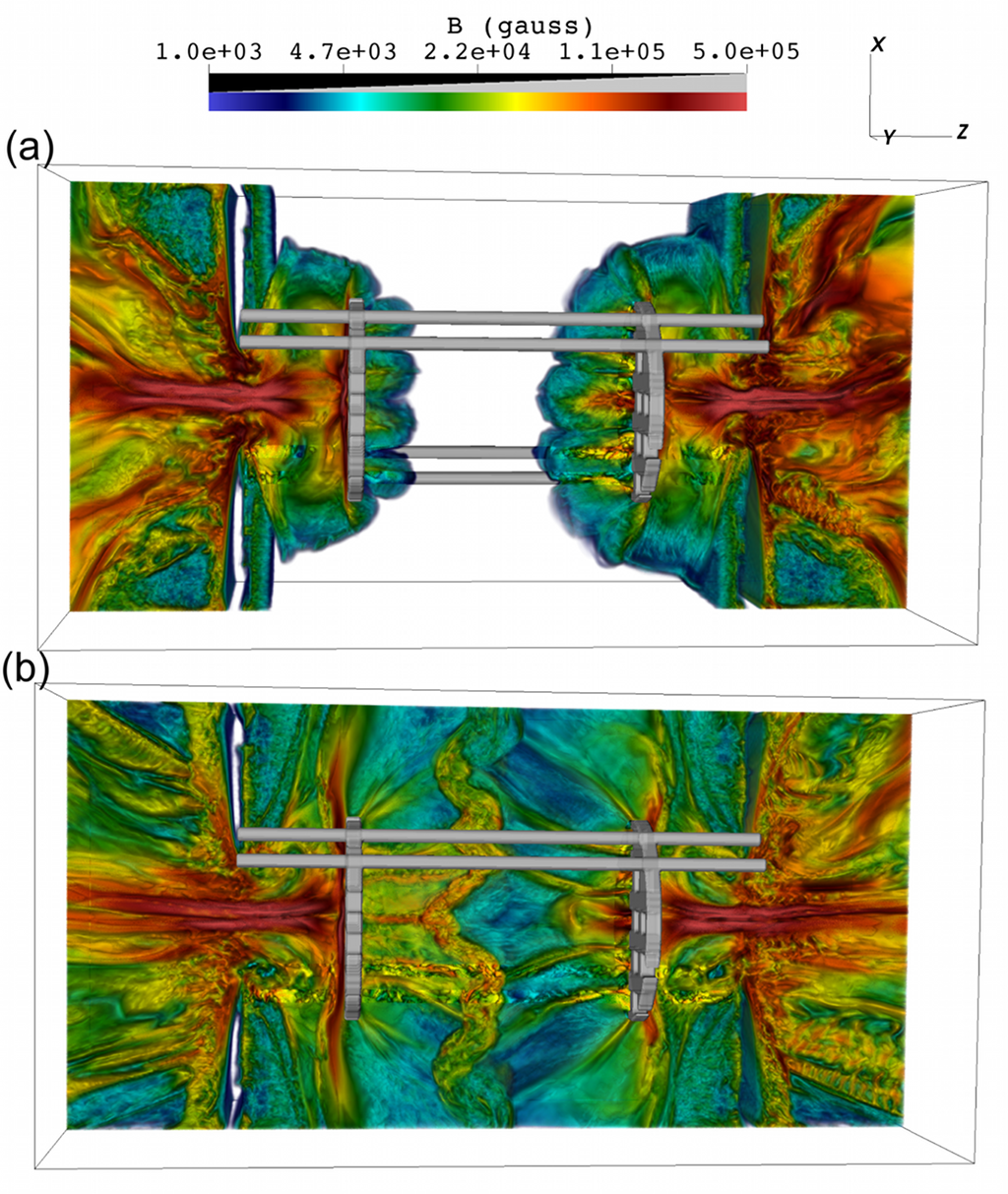}
 \caption{Volume rendering of the magnetic field magnitude in Gauss for the 1Cl10ns case at (a) 20 ns and (b) 35 ns.}
\label{fig:Field}
\end{figure}

\begin{figure*}[t]
\begin{center}
\includegraphics[width=6.5in]{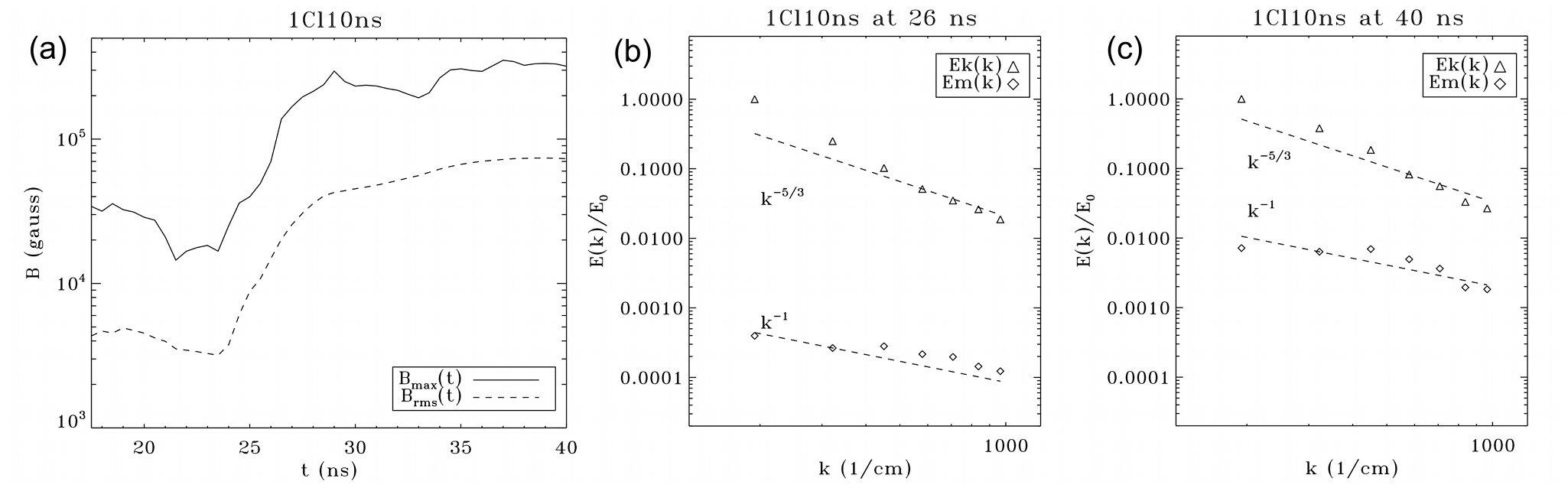}
\caption{(a) Magnetic field strength (maximum value and root mean square value) as a function of time in the tracking
control volume. (b) Angle integrated kinetic ($E_k$) and magnetic energy ($E_m$) as a function of k, for case 1Cl10ns at t $=$ 26 ns. The
values are normalized to the kinetic energy at the largest scale and we display a scale range between 50 and 500 $\mu$m.
Immediately after collision the magnetic energy is only a small fraction of the kinetic energy. The magnetic energy
follows a power law consistent with k$^{-1}$ while the kinetic energy displays a Kolmogorov slope. At the largest scales there is a
steepening due to bulk motion in the Z direction.
(c) Same as (b) but for t $=$ 40 ns. At saturation the magnetic energy becomes comparable to the kinetic energy ($1-10\%$).
The kinetic energy slope-steepening at large scales is less pronounced as the turbulence homogenizes more.}
\label{fig:Spectra}
\end{center}
\end{figure*}

\section{Discussion}
\label{Discussion}

\subsection{Predicted plasma properties}
\label{label:plasma}

The general behavior described in Figure \ref{fig:DensEvolution} occurs in all four simulations,
but the change in composition and drive affect both the timing of events and the plasma properties.
This becomes apparent when comparing the front positions at the same evolution time.
In Figure \ref{fig:timing}) we display a half-rendering of the electron density logarithm at 20 ns for all cases; the 6\% chlorine-doped
cases (panels a and b) appear slower than the 1\% cases (panels c and d). This is a consequence of variation in the
foil target density, which is 20\% larger for cases 6Cl10ns and 6Cl5ns. Similarly, the decrease in drive duration from
10 to 5 ns directly translates to an increase of laser intensity, which results in faster flows for cases 6C5ns and 1Cl5ns 
(panels b and d) than their 10 ns counterparts (panels a and c). In all cases, however, the flows eventually collide to form a
turbulent interaction region, reaching high temperatures that endure for several nanoseconds. Thus, a natural separation in terms
of analyzing the simulation results is to consider the flows prior to collision and the turbulent region after the collision,
circumventing the temporal offsets due to different drives and compositions. 

To present quantitative results, we utilize a control volume, a cubic box of edge length 500 $\mu$m, to sample relevant
plasma quantities before and after collision. In the former case, the box tracks in time the propagating plasma front from grid A,
centered at the edge of the front. Post-collision, the tracking volume is pinned at the interaction region, centered at the
stagnation point formed by the colliding fronts. The box is allowed to move along the line of centers (LoC) that is parallel to
the $Z$ axis and intersects the centers of the targets ($X=Y=0$). A comprehensive list of plasma properties for case 1Cl10ns is given 
in Table \ref{table:PlasmaProperties}. Similar values are also recovered for the remaining cases with some variation due to the drive
and composition difference.  

The plasma remains highly collisional throughout the simulation and the MHD treatment is valid; the distribution
function can be approximated with a Maxwellian\cite{Tzeferacos2017}. Prior to collision, the flows are mildly supersonic ($M\sim2-3$).
As the plasma flows traverse the grids, weak shocks are formed that result in the heating of ions, whereas electron temperature lags
slightly behind due to the initially long ion-electron equilibration timescale. Typical flow densities and temperatures are of the order of
$\sim 10^{17}-10^{18}\,\textrm{cm}^{-3}$ and few tens of eV, respectively. The flows propagate with velocities of a couple of hundred
km s$^{-1}$ to meet at the domain center. From the laser-target interaction, we have the generation of strong
Biermann battery\cite{Biermann1950} magnetic fields, which are of the order of $\sim$ MG close to the targets and are advected with the
plasma. The misaligned gradients of electron pressure and density continuously generate fields as the flows propagate but advection
causes substantial spatial dilution, reducing the field strength of the fronts down to values of $\sim 1-10$ kG prior to
collision (Figure \ref{fig:Field}a).  

The collision takes place at $\sim 24-25$ ns for the 1Cl10ns case and results in a pair of
accretion shocks with a subsonic turbulent region in between. The ion and electron temperatures increase to a few hundreds of eV
(Figure \ref{fig:TeleRm}a) and equilibrate rapidly. While such turbulent flows were recreated also in our simulations of the
colliding jets experiment\cite{meinecke2015} with the Vulcan laser, in the simulations of the Omega platform we reach values of Rm in the
many hundreds (Figure \ref{fig:TeleRm}b). The high Rm values persist for several ns after the collision (Figure \ref{fig:TeleRm}b) and
the magnetic fields appear significantly amplified to peak values of hundreds of kG (Figure \ref{fig:Field}b).  

In the simulations, the turbulent plasma is characterized by an outer scale $\cal{L}\sim$ 600 $\mu$m and has a Kolmogorov-like spectrum (Figure \ref{fig:Spectra}).
The dissipation scales are below our spatial resolution, both for viscous ($l_{\nu}= \cal{L}$/Re$^{3/4}\sim 1\,\mu\textrm{m}$)
and resistive ($l_{\eta} = \cal{L}$/Rm$^{3/4} \sim 4\,\mu\textrm{m}$, for $\textrm{Pm}<1$) dissipation. As a result, the simulations cannot capture the complete energy cascade
but can inform us on the behavior of the energy spectra at larger scales -- in the limited range allowed by our numerical resolution. Using
the control volume mentioned above, we can recover the temporal evolution of the magnetic field strength and compute, at different times,
the angle-integrated spectra of the magnetic and kinetic energy. The simulated time history of the field (peak values $B_{\textrm{max}}$ and root mean
square values $B_{\textrm{rms}}$ in the control volume) for case 1Cl10ns is given in Figure \ref{fig:Spectra}a. This semi-log plot
shows the sequence of events: initially ($t<24$ ns), the magnetic field decreases as the plasma expands (a dilution phase) to values of a few kG,
which will act as seed fields for the dynamo amplification; then collision occurs ($t\sim24-25$ ns), and we see a sharp increase due to compression
effects and an exponential increase phase ($ t \sim 25-28$ ns) consistent with kinematic dynamo;
subsequently ($t>28$ ns), the exponential growth phase ends as the field strength increases, entering a non-linear dynamo phase where the field becomes important with respect to the flow dynamics\cite{Schekochihin2002};
the curve flattens at later times when saturation is reached, with peaks as high as
$\sim 300-350$ kG. This occurs on timescales that are comparable to an eddy turnover time at the outer scale,
$t_{{\cal{L}}} \sim {\cal{L}}/{u} \sim$ 4 ns.
Panels b and c of Figure \ref{fig:Spectra} show the simulated spectra for the one-dimensional, angle-integrated kinetic and magnetic energies,
\begin{equation}
\DS E_k(k) = \frac{1}{2}\left< \rho\right>\int d\Omega_k k^2 \left< | {\bf u(k)} |^2 \right>\,\textrm{and}
\end{equation}
\begin{equation}
\DS E_m(k) = \frac{1}{2}\int d\Omega_k k^2 \left< | {\bf B(k)} |^2 \right>,
\end{equation}
at different times -- shortly after collision on panel b and at saturation in panel c. The kinetic energy follows a $k^{-5/3}$
Kolmogorov power law, consistent with a subsonic turbulent plasma. The magnetic energy, on the other hand, follows
a $k^{-1}$ power law, previously found for galactic turbulence \cite{Ruzmaikin1982} and fluctuation dynamo at small magnetic Prandtl
numbers\cite{schekochihin2007}. Shortly after collision (Figure \ref{fig:Spectra}b) the magnetic energy is considerably
smaller than the kinetic energy. At saturation (Figure \ref{fig:Spectra}b) the magnetic energy rises up to 1-10\% of the
kinetic energy, depending on scale. Such saturation values were also recovered by other numerical studies of turbulent
dynamo\cite{Schekochihin2004,Federrath2011}.

As a whole, our numerical results suggest that the Omega laser experiments that we have simulated would be able to reach Rm close to critical
values\cite{schekochihin2007} and thus enter the turbulent dynamo regime, to enable experimental study of the properties of MHD turbulence and
magnetic field amplification. For a discussion on the experimental findings the reader is refered to our companion article\cite{Tzeferacos2017}.

\subsection{Validation of the simulations}
\label{label:validation}

During the experiment we fielded a number of diagnostics to probe the plasma and magnetic field properties\cite{Tzeferacos2017}.
A small subset of the experimental data can be used to validate specific properties of the simulations,
such as the propagation speed of the colliding flows and the time of collision timing. For these we utilize information from
soft X-ray imaging and the Thomson scattering diagnostic\cite{Evans1969}. 

\begin{figure}
\includegraphics[width=\columnwidth]{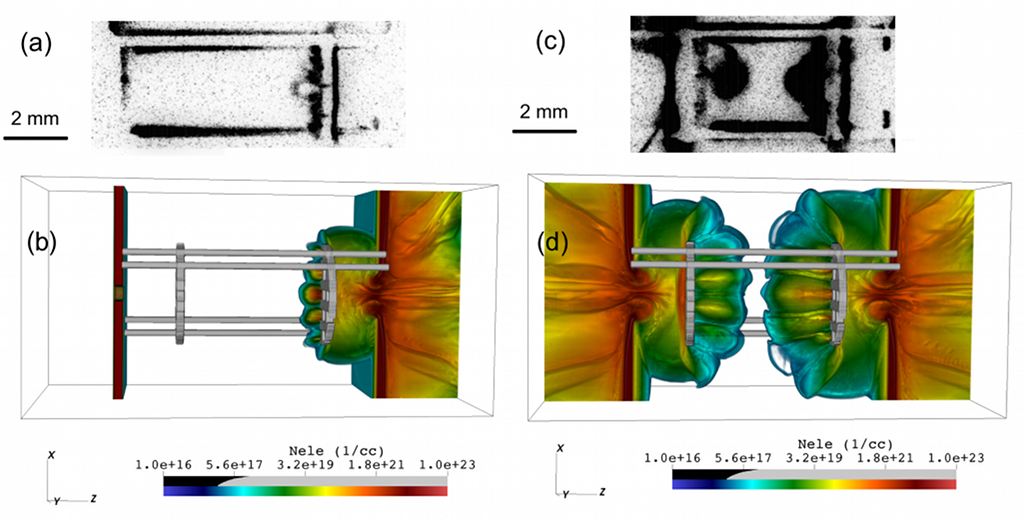}
 \caption{Soft X-ray images of the plasma fronts at two different times, (a) 24 ns and (c) 31 ns. The experimental configuration
 corresponds to our simulation case 6Cl10ns. (b) Rendering of the electron density logarithm for our 6Cl10ns simulation at 24 ns,
 omitting the plasma from grid B. (d) Same as (b) at 31 ns.}
\label{fig:Fronts}
\end{figure}

\begin{figure}
\includegraphics[width=\columnwidth]{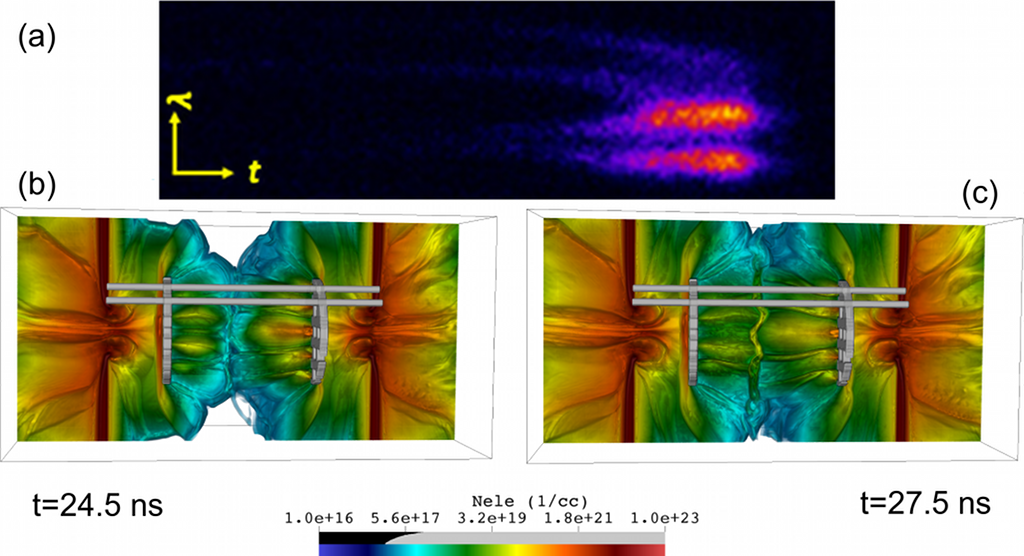}
 \caption{Flow collision timing from the Thomson scattering diagnostic for an experimental shot that corresponds to our 1Cl10ns case.
(a) Time-streaked image of the Thomson scattered light, with a temporal resolution of $\sim 50$ ps. The four peaks correspond
to a pair of ion features that move in opposite directions and merge at late times as a single-flow plasma forms.
The image corresponds to the $24.5-27.5$ ns time interval, indicating that collision occurs between $25-26$ ns.
(b) Volume rendering of the electron number density logarithm at 24.5 ns for case 1Cl10ns. The counter-streaming flows reach the probing region.
(c) Same as (b) at 27.5 ns. The turbulent interaction region is well-formed.}
\label{fig:Collision}
\end{figure}

Experimental X-ray images taken at early times of the evolution allow us to track the position of the plasma fronts prior to collision. 
These are given in Figure \ref{fig:Fronts}, panels a and c. The experimental configurations correspond to our 6Cl10ns case at
24 and 31 ns, respectively. Notice that in the case of panel a, we only drove the target on the side of grid A and
grid B was missing from the assembly. This can be reproduced with our 6Cl10ns simulation by omitting the evolution from the side
of grid B. The agreement between experimental data and simulation results is fairly good (Figure \ref{fig:Fronts}, panels b and d).
The heavily chlorinated target propagates slowly -- slower than the rest; see also Figure \ref{fig:timing} -- and at 24 ns the
front has just crossed grid A. At 31 ns, the two flows are clearly visible and have yet to meet at the center of the chamber. It should
be noted that the X-ray emission depends on the plasma density, therefore the panel c only shows emission from the denser parts of the
flow. The image is unfortunately saturated due to diagnostic filter options and we cannot discern variations in the plasma structure.
To bound the collision timing and validate the numerically predicted time, we can utilize information from the Thomson scattering diagnostic.
The spectrum of light produced by Thomson scattering in a hot plasma, in particular the shape and position of the ion feature that results from collective
processes involving excitation of ion-acoustic wave modes, depends sensitively on the plasma velocity, electron density, electron temperature, and ion temperature.
A scattering diagnostic based on this effect thus allows detailed inferences of these physical quantities\cite{Evans1969}. 
A 2$\omega$ low energy beam is focused on a small spot in the plasma; a dedicated detector records the radiation from a
narrow angle range to produce a time-streaked image of the scattered light from the small target volume.
In one of the shots, corresponding to our 1Cl10ns case, the diagnostic probed the interaction region in the time interval $24.5-27.5$ ns, which
overlaps with the numerically predicted collision time ($\sim 24-25$ ns). While a typical spectrum would exhibit only one
pair of ion features (two peaks in the intensity profile), in this case we observed four peaks (Figure \ref{fig:Collision}a).
This occurs when the light scatters off counter-streaming plasma, i.e., when the plasma fronts converge. At later times, the four peaks merge
into two, an indication that the interaction region has formed in the experiment. This sequence and timing of events matches fairly
closely the simulation results, within $\sim 1-2$ ns (panels b and c of Figure \ref{fig:Collision}). It should be noted that this agreement
was achieved \emph{without} tuning of the laser energy deposition, which can sometimes be necessary to account for laser-plasma interaction
(LPI) effects that can reduce the drive efficiency. 

\subsection{Comparison of simulated diagnostics and experimental data}

\label{label:diagnostics}
\begin{figure}
\includegraphics[width=\columnwidth]{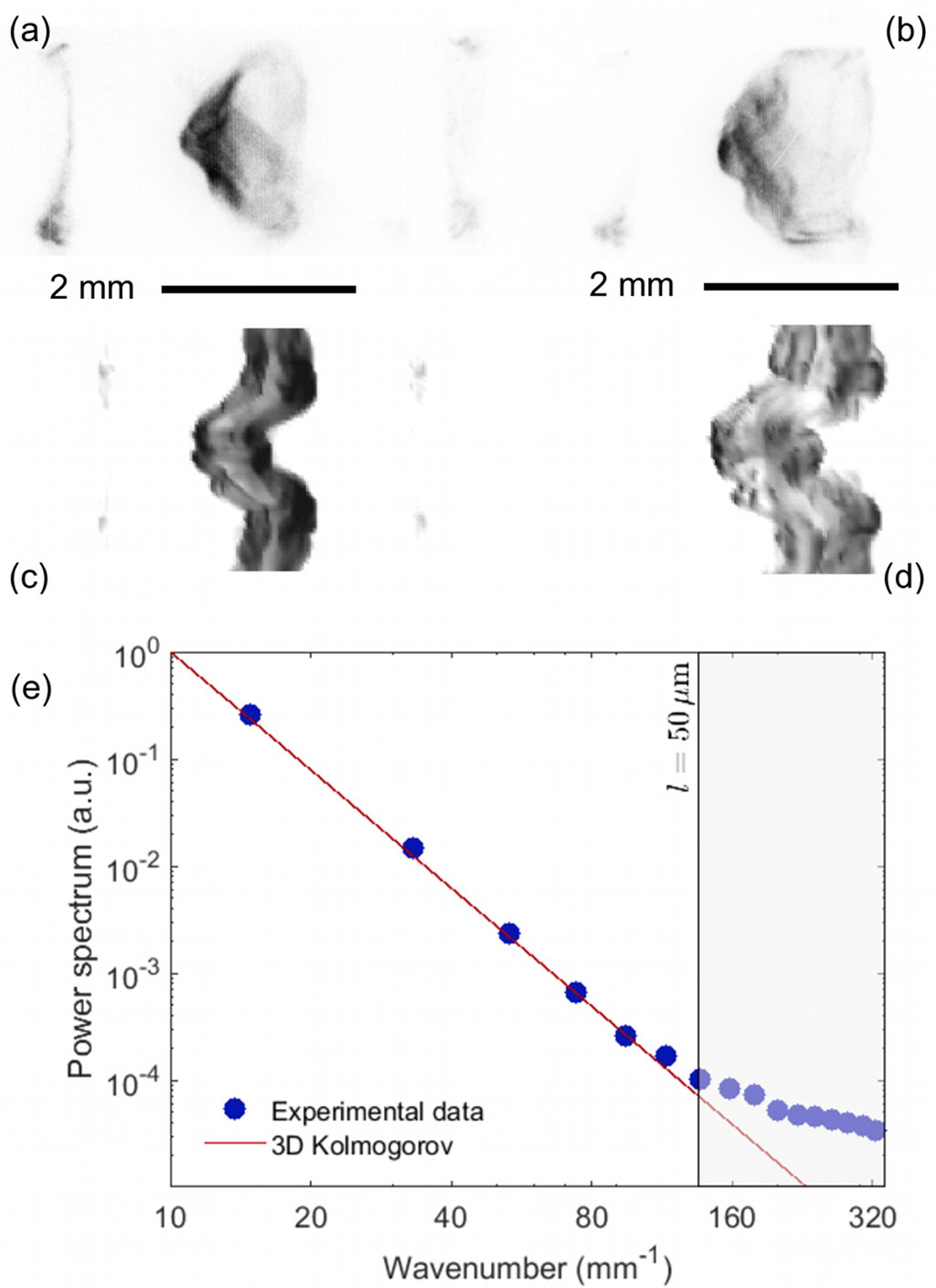}
 \caption{Experimental and synthetic X-ray images and power spectrum. (a) Soft X-ray experimental image at 27 ns for a shot corresponding to our
 1Cl5ns case. (b) Same as (a) but at 31 ns. (c) Synthetic X-ray image from the FLASH results for case 1Cl5ns at 27 ns. (d) Same as (c) but at 31 ns.
 (e) Power spectrum recovered from the spectral analysis of the interaction region in panel (b). The power spectrum of 2D intensity fluctuations
 is proportional to the 3D spectrum of density fluctuations\cite{churazov2012,Tzeferacos2017} and is consistent with Kolmogorov turbulence.
 The deviation seen in small scales is attributed to Poisson noise.}
\label{fig:Xray}
\end{figure}

There has been considerable effort in the FLASH code development to implement synthetic counterparts of experimental diagnostics
that are commonly used by the HEDLP community. This reflects an effort to cast simulation results in a format that
allows them to be compared \emph{directly} to the data, minimizing post-processing of the latter and including as
many of the physical and statistical processes that go into the creation of the experimental image. Here we consider
three of the diagnostics that were fielded in our Omega experiment\cite{Tzeferacos2017}, used to probe the state of the
plasma and the magnetic field in the interaction region. 

{\bf X-ray imaging.} The X-ray imaging, which was used above to validate the FLASH predictions of the propagation speed of the plasma fronts,
can provide useful information regarding the shape and properties of the interaction region. To create synthetic images from the FLASH
simulations, the results were recast to be read in post-processing by SPECT3d\cite{macfarlane2003}, a collisional-radiative spectral
analysis code designed to simulate atomic and radiative properties of laboratory plasmas. We create synthetic images for case 1Cl5ns
at 27 and 31 ns and compare them with the experimental recovered X-ray images (Figure \ref{fig:Xray}). In the experimental results,
we can see that the interaction region is well-formed by 27 ns (panel a), with the characteristic ``cup'' shape
predicted by our numerical models (we remind the reader that case 1Cl5ns was the one that exhibited flow collision very early on at $\sim 20-21$ ns,
see also Figure \ref{fig:timing}). At 31 ns, the interaction region has become slightly thicker 
(panel b). The FLASH/SPECT3d synthetic images are in good agreement with the overall shape and distinct features of the interaction region
(panels c and d), exhibiting also the same trend in the thickness. 
The turbulence also has a measurable effect on X-ray emissivity. The 2D fluctuations in X-ray intensity can be related to the
3D density fluctuations\cite{churazov2012} and, under specific caveats which include negligible electron temperature fluctuations and
isotropic turbulence, have proportional power laws. A formal discussion on this proportionality and on X-ray image analysis is
presented in the companion paper\cite{Tzeferacos2017}. The spectral analysis for the 31 ns experimental image is consistent with
a 3D Kolmogorov power law $k^{-11/3}$ (Figure \ref{fig:Xray}e); this corresponds to a 1D power spectrum
$\propto k^2 k^{-11/3} \sim k^{-5/3}$. If the caveats apply and the interaction region in the experiment is indeed subsonic,
as the simulations seem to indicate, then the density would behave as a passive scalar and
the kinetic energy power spectrum would also follow a $k^{-5/3}$ power law. Such a result would agree with the FLASH
prediction (panels b and c of Figure \ref{fig:Spectra}). 

{\bf Thomson scattering.} As we mentioned above, the Thomson scattering diagnostic probes the plasma with a low-energy 2$\omega$ laser beam (526.5 nm), to produce a light
spectrum with features sensitive to the plasma characteristics\cite{Evans1969}. We have implemented in FLASH a simulated Thomson
scattering diagnostic to reproduce such spectra. The code module computes multiple ray paths, each going from a lens
location to a scattering location, and then from the scattering area to the detector (alongside the diagnostic rays, we also
launch rays from our laser package to account for any laser heating effects\cite{Tzeferacos2017}). While the lens center
and the detector location are held fixed, multiple ray paths are generated by iteration over points in the part of the region of
interest, at subcell resolution.  We perform an integration along the paths to compute the attenuation of ray
power by inverse bremsstrahlung; the simulated diagnostic thus takes into consideration the effect of matter present in
the chamber on both the incoming and the scattered light via absorption. Each ray determines a scattering angle and plasma state
(electron density, electron/ion temperature, bulk velocity components, and turbulent velocity\cite{Inogamov2003, meinecke2015});
from these characteristics, a Thomson spectrum is computed for each ray using the approximations and code developed by Froula and coauthors\cite{Froula2010}. The overall
simulated sum is then a weighted sum of the contributions from rays, where the weights include the effects of probe beam shape and attenuation.

\begin{figure}
\includegraphics[width=\columnwidth]{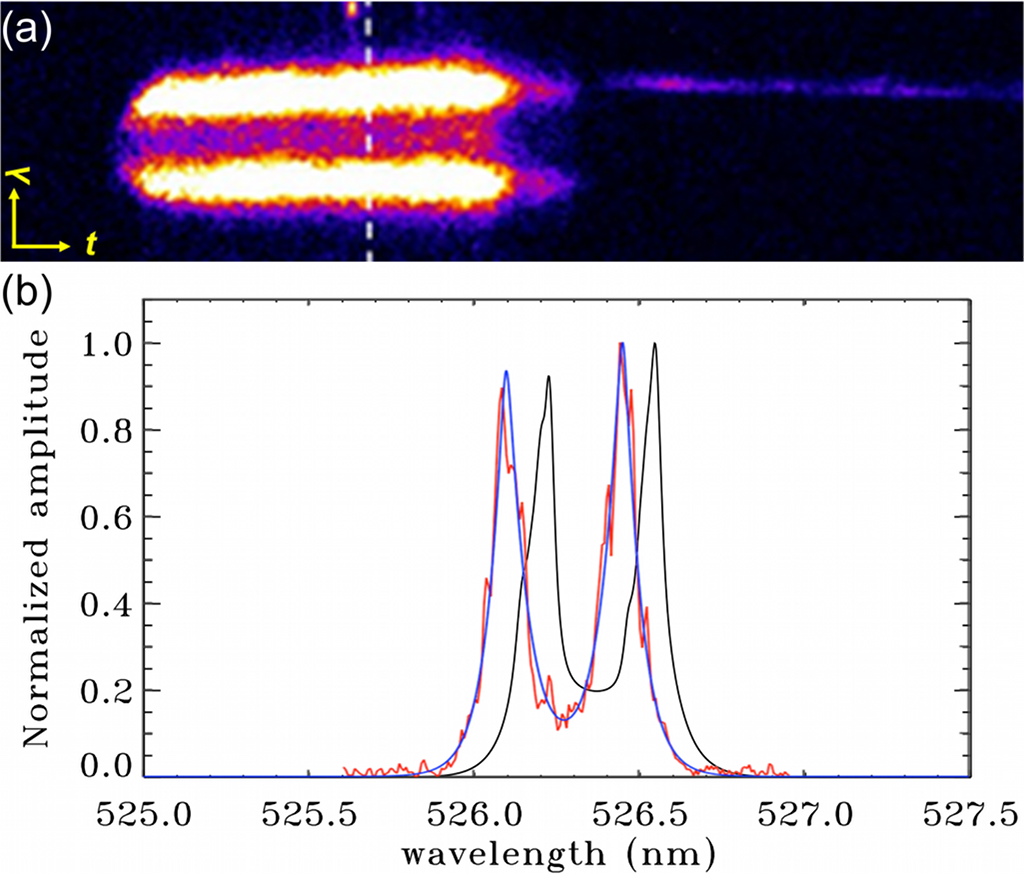}
 \caption{(a) Time-streaked image of the Thomson scattered light, for an experiment corresponding to our 1Cl10ns case. The probe beam is
 on for a 1 ns interval between 32.5 and 33.5 ns, targeted at the interaction region. The two ion features can be
 used to characterize the plasma properties\cite{Evans1969}. The white dashed line denotes the locus where the frequency lineout in panel (b) was
 taken. (b) Intensity profile at $t \sim 33$ ns. The red line is the experimental data, the blue line is an analytic fit without instrument
 noise, and the black line is the FLASH prediction. The shape of the ion features is in fairly good agreement with the experimental data,
 albeit there is a frequency offset between the features. This could be explained if the bulk velocity along the LoC found
 with FLASH were smaller by $\sim 50$ km s$^{-1}$ with respect to the one in the experimental data.}
\label{fig:TS}
\end{figure}

The experimental data for a shot corresponding to our 1Cl10ns simulation is shown in Figure \ref{fig:TS}a. The Thomson scattering
laser is on for 1 ns and probes a small ($\sim 50$ $\mu$m focal spot) volume between 32.5 and 33.5 ns. The pair of ion-acoustic features
is clearly visible, along with a stray-light line at the laser wavelength (265.5 nm). The white dotted line denotes the locus where we extract
a wavelength lineout to analyze the features (Figure \ref{fig:TS}b). The red line is the experimental data, the blue line is
an analytic fit without instrument noise, and the black line is the FLASH prediction from the synthetic Thomson scattering diagnostic.
The simulated spectrum agrees fairly well with the experimental result, in terms of shape, separation, and width
of the ion features; these characteristics depend on the plasma properties discussed in \S \ref{label:plasma}.
The discrepancy in terms of position -- which is defined by Doppler shift due to the bulk velocity in the scattering
volume -- could be explained if the plasma in the experiment had a bulk velocity component along the LoC that
is $\sim 50$ km s$^{-1}$ larger than the simulation. It is worth noting here that, due to the small volume probed by
the Thomson scattering diagnostic, our correct prediction of the shape of the interaction region was crucial:
had we focused the laser beam at the center of the target chamber -- where we would expect collision to occur -- we would have missed the
interaction region.

{\bf Proton radiography.} To measure and characterize the magnetic field in the plasma, we use monoenergetic proton radiography\cite{Li2006}. 
This experimental diagnostic technique images magnetic fields using proton emission from the laser-driven implosion of a small D$^3$He capsule.
The capsule is located 1 cm away from the target chamber center (the center of our computational domain) and its implosion causes a quasi-isotropic emission
of protons at $\sim 3$ and $\sim14.7$ MeV. The protons traverse the interaction region and interact only with the magnetic
fields -- as other physical effects such as collisions or kinetic effects are negligibly small. The protons are subsequently
recorded on a detector (a CR39 plate) 28 cm away from the capsule.
The deflection of a proton's path bears information on the morphology and strength of the magnetic fields that caused it;
from the two-dimensional image we can infer the path-integrated magnetic field\cite{Graziani2016,Bott2017}, provided that the
fields are not too strong.

We have implemented in FLASH a proton radiography synthetic diagnostic. The module fires protons towards the simulation domain and records
their deflection due to electric and magnetic fields on a detector screen. By employing conical beams the code can efficiently emulate a spherical
sector of the isotropic emission, reducing considerably the computational cost of treating billions of protons -- a typical proton yield of such a
capsule implosion. Each proton is traced separately and is initialized with random velocity vectors in a spherical volume equal to the size of the
capsule at bang time ($\sim 40$ $\mu$m). The protons' deflections are calculated using the Lorentz force, assuming the electric and magnetic fields
do not change during the traversal of the domain by the protons. For each cell in the domain, the electric and magnetic fields are averaged from
their staggered representations\cite{Lee2013} and they are considered constant within each cell. The protons are collected on a screen, where we
record their final position on the screen's coordinate system.   

\begin{figure*}[t]
\includegraphics[width=6.5in]{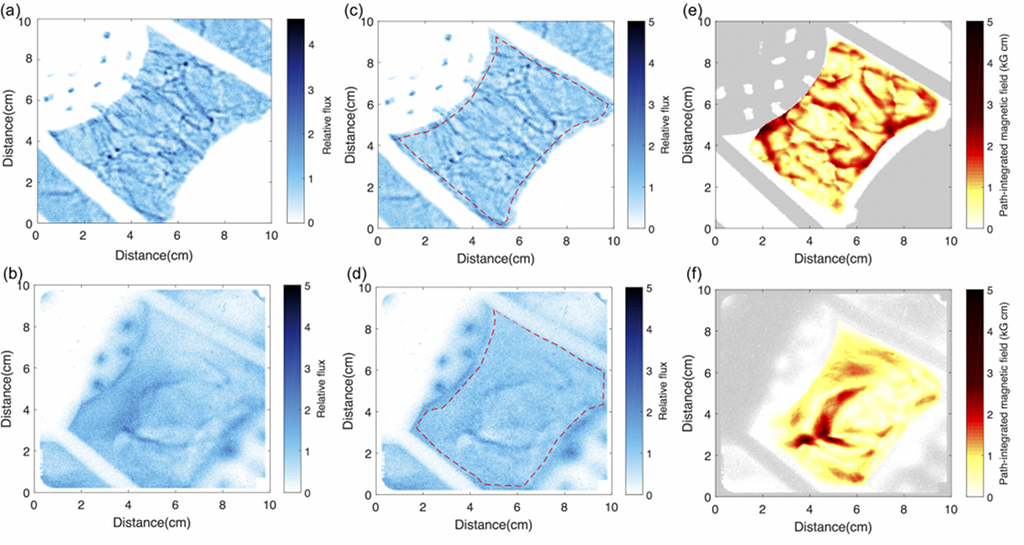}
 \caption{Proton radiography images and path-integrated magnetic field reconstruction.
 (a) Simulated proton radiograph for case 1Cl10ns at 31 ns. Shown is the relative proton flux for the 14.7 MeV protons.
 The protons are obstructed by grids and rods, which are clearly visible. Grid A lies on the top left
 and grid B on the bottom right. The filamentary structures are the result of the magnetized turbulence that we have in the simulation.
 (b) Proton radiograph from an experimental configuration that corresponds to the case shown in panel (a). While a few filamentary structures
 exist, the experimental image exhibits more smearing, less small-scale structure, and systematic large-lengthscale variation
 in the proton flux\cite{Manuel2012}. (c) Same as (a), but with a low-pass filter applied in the area
 denoted by the red dashed line, to remove the large-lengthscale variations. Since the synthetic image was made with an isotropic
 proton beam, the filtering has no effect. (d) Same as (b), but with a low-pass filter applied as in (c). The filamentary structures
 remain unaffected by the filtering. (e) Path-integrated magnetic field reconstruction on the synthetic proton radiograph shown in (c).
 (f) Same as (e), but for the experimental image in (d).}
 \label{fig:PR}
\end{figure*}

The simulated proton radiograph for case 1Cl10ns at 31 ns is shown in Figure \ref{fig:PR}a for the 14.7 MeV protons.
The filamentary structure seen between the grids (grid A top left, grid B bottom right) is the result of proton deflection by
the magnetic fields that develop in the turbulent interaction region, in the simulation. The synthetic image includes a number of smearing effects
present in experimental images, such as the smearing due to the finite capsule size, the binning of protons, and Poisson
noise\cite{Graziani2016}. However, this list is not exhaustive and other plasma and instrument effects may affect the experimental proton radiograph.
Experimental images suffer also from long lengthscale variations $\gtrsim 50\%$ in the
proton flux that should be taken into consideration in a quantitative analysis\cite{Manuel2012}. 
The experimental proton radiograph that corresponds to the 1Cl10ns case is shown in Figure \ref{fig:PR}b.
This particular image shows only a few filamentary imprints on the CR39 plate. Moreover, it has more pronounced smearing
and lacks the small-scale structure that we see in the simulated radiograph.

To evaluate quantitatively the field strength and topology from the experimental image and compare to the synthetic radiograph,
we can apply either linear\cite{Graziani2016} or non-linear\cite{Bott2017} reconstruction techniques
to determine path-integrated fields. Since we are in the order-unity contrast regime, we utilize the latter. The first step in this
analysis is to apply a low-pass filter on the proton radiographs to remove systematic
large lengthscale variations\cite{Manuel2012}, to which non-linear reconstruction techniques are sensitive (panels c and d of Figure \ref{fig:PR}).
This is performed on the areas denoted by the red dashed lines. The synthetic image remains unaltered due to the assumption
of isotropic proton emission in the FLASH code, and the experimental image retains all its original sharp structures. Next
we apply the reconstruction to recover path-integrated magnetic field values in the image plane (panels e and f of Figure \ref{fig:PR}).
The number of path-integrated magnetic field structures is smaller in the experimental image, as expected following from the
reduced number of filaments in the radiograph. We do however find agreement between the path-integrated magnetic field values
($\sim 3-5$ kG cm), which in turn can yield estimates of the magnetic field strength in the interaction region\cite{Tzeferacos2017}.

\section{Conclusions}
\label{Conclusions}

The generation and amplification of magnetic fields observed in the universe is an ongoing research topic of modern astrophysics.
While a number of mechanisms have been proposed to address generation of seed magnetic fields\cite{Kulsrud1997, schlickeiser2003},
the amplification is primarily attributed to turbulent dynamo\cite{Ryu16052008,schekochihin2006,subramanian2006}, where the stochastic
motions of the turbulent plasma can stretch and fold seed magnetic fields, amplifying them until they become dynamically
significant\cite{Schekochihin2002}. While a significant body of theoretical work exists on this process, turbulent dynamo
has eluded systematic study in the laboratory due to the large magnetic Reynolds numbers it requires to operate,
especially in the small Pm regime\cite{schekochihin2007}.

In this article, we described the numerical effort to design an experiment that could enable us to reach the turbulent dynamo 
regime. The results presented here highlight the advantages of using numerical modeling for experiment design and analysis.
When combined with synthetic diagnostics, simulations can predict expected signals and be a crucial guide in determining the
placement and timing of experimental diagnostics. Validated simulations produce data that can be analyzed quantitatively,
allowing strong conclusions to be drawn from them.
 
The design of the experimental platform was based on our previous work (Figure \ref{fig:vulcan}) on laser-driven
plasmas\cite{Gregori2012,meinecke2014,meinecke2015} and the simulations described here.
Our simulation campaign employed all the recently-developed HEDLP capabilities of the FLASH code
and ANL's {\em Mira} BG/Q supercomputer. The configuration (Figure \ref{fig:Target}) that we designed consists of two
diametrically opposed targets that are backlit with temporally stacked beams, which deliver 5 kJ of energy on each side;
the beams drive a pair of colliding plasma flows that carry seed magnetic fields generated via Biermann battery and
propagate through a pair of grids, which destabilize them, introducing a driving scale; the flows meet at the center of the chamber to form a hot,
turbulent interaction region (Figure \ref{fig:DensEvolution}), where we measure the plasma properties (Table \ref{table:PlasmaProperties}).

In the simulations, the turbulent plasma achieves sufficiently large Rm values that dynamo can act on the small
seed fields and amplify them by a factor of $\sim 25$, reaching saturation within $1-2$ eddy turnover times at the
outer scale (Figures \ref{fig:TeleRm}, \ref{fig:Field}, and \ref{fig:Spectra}).
The peak field values are of the order of $\sim 300-350$ kG, with a magnetic-to-kinetic
energy ratio of $\sim 1-10$ \%, depending on the scale considered. In the modest dynamic range that we have in the simulations, 
the kinetic energy shows a Kolmogorov-like $k^{-5/3}$ power spectrum and the magnetic energy shows a $k^{-1}$ power spectrum,
which are consistent with dynamo. This result provides upper bounds for the critical Rm value required by dynamo to operate:  
for the Re values achieved in the simulations, Rm$_c \lesssim 900-1300$, consistent with the constraints derived for the small Pm
regime\cite{schekochihin2007}.

The FLASH simulations were validated against a small subset of experimental data from our Omega experiments\cite{Tzeferacos2017};
we found good agreement in the propagation speed of the colliding fronts (Figure \ref{fig:Fronts})
and the collision timing (Figure \ref{fig:Collision}). Moreover, the development of simulated diagnostics
allowed us to compare synthetic vs. experimental data from the interaction region, and 
find good agreement in terms of the shape of the interaction region (Figure \ref{fig:Xray}) and the plasma properties (Figure \ref{fig:TS}).
Nevertheless, some comparisons remain inconclusive: while FLASH predictions of a subsonic Kolmogorov MHD turbulence are
consistent with the density power spectrum recovered from the X-rays, the filamentary structures seen in proton radiography
are dissimilar, despite the apparent agreement in path-integrated magnetic field strength (Figure \ref{fig:PR}). To go beyond these
validation-geared comparisons, a full analysis of the experimental dataset is needed;
even if the simulation results indicate that the experimental platform is capable of demonstrating turbulent dynamo,
the final outcome can only be decided by our experimental results, discussed in a companion paper\cite{Tzeferacos2017}.

\section*{Acknowledgments}

We thank the anonymous reviewer whose constructive input improved the article's presentation.
This work was supported in part at the University of Chicago
by the U.S. Department of Energy (DOE) under contract B523820
to the NNSA  ASC/Alliances Center for Astrophysical Thermonuclear
Flashes; the U.S. DOE NNSA ASC through the Argonne Institute for Computing in Science under field
work proposal 57789; the U.S. DOE NNSA through the NLUF grant
No. DE-NA0002724; the U.S. DOE Office of Science through grant No. DE-SC0016566; and
the U.S. National Science Foundation under grants PHY-0903997 and PHY-1619573.
Support from AWE plc., the Engineering and Physical Sciences
Research Council (grant numbers EP/M022331/1; EP/N014472/1; and EP/N002644/1) and the
Science and Technology Facilities Council of the United Kingdom is acknowledged.
This work was supported in part by the National Research Foundation of Korea through grant 2016R1A5A1013277.
Awards of computer time were provided by the DOE Innovative and Novel
Computational Impact on Theory and Experiment (INCITE) and ASCR Leadership
Computing Challenge (ALCC) programs. This research used resources of the Argonne Leadership Computing
Facility at Argonne National Laboratory, which is supported by the
Office of Science of the U.S. Department of Energy under contract DE-AC02-06CH11357.
The research and materials incorporated in this work were partially developed at the National
Laser Users’ Facility at the University of Rochester’s Laboratory for Laser Energetics, with
financial support from the U.S. Department of Energy under Cooperative Agreement DE-NA0001944.

%\nocite{*}
\bibliography{ms}% Produces the bibliography via BibTeX.

\end{document}